\documentclass[10pt]{ws-rv975x65}
\usepackage{ws-rv-van}
\usepackage{latexsym}
\usepackage{amssymb}
\usepackage{amsbsy}
\usepackage{amsmath}
\usepackage[varg]{txfonts}
\usepackage{mathrsfs}
\usepackage{upgreek}
\usepackage{verbatim}
\usepackage{array}
\usepackage{color}
\usepackage{graphicx}
\usepackage{hyperref}
\DeclareMathAlphabet{\mathpzc}{OT1}{pzc}{m}{it}
\title{Pairing correlations with single Cooper pair transfer to individual quantal states}
\begin{document}
\chapter{Pairing correlations with single Cooper pair transfer to individual quantal states}
\author[G. Potel]{G. Potel}
\address{Departamento de Fisica Atomica, Molecular y Nuclear, Universidad de Sevilla, Facultad de Fisica, Avda. Reina Mercedes s/n, Spain.}
\author[R. A. Broglia]{R. A. Broglia}
\address{Dipartimento di Fisica, Universit\`{a} di Milano,
Via Celoria 16, 20133 Milano, Italy.}
\address{INFN, Sezione di Milano Via Celoria 16, 20133 Milano, Italy.}
\address{The Niels Bohr Institute, University of Copenhagen, Blegdamsvej 17,
2100 Copenhagen {\O}, Denmark.}
\body
\begin{abstract}
Making use of the fact that the collective modes associated with the spontaneous (static and dynamic) violation of gauge invariance in atomic nuclei (pairing rotations and pairing vibrations), are amenable to a simple, quite accurate nuclear structure description (BCS and QRPA respectively), it is possible to quantitatively test the reaction mechanism which is at the basis of two--nucleon transfer reactions, specific probe of  pairing in nuclei. With the help of the static and dynamic mean field spectroscopic amplitudes, taking into account successive and simultaneous transfer channels properly corrected because of non--orthogonality effects, as well as describing the associated elastic channels in terms of experimentally determined optical potentials, one obtains absolute, two--particle transfer differential cross sections which provide an overall account of the data within experimental errors. One of the first results connected with such quantitative studies of pairing correlations in nuclei is the observation of phonon mediated pairing in the exotic halo nucleus $^{11}$Li, and the associated discovery of a new mechanism to break nuclear gauge symmetry: bootstrap, pigmy--resonance--mediated Cooper pair binding. 
\end{abstract}
\section{Gauge invariant BCS nuclear theory:Pairing vibrations and pairing rotations}
A closed shell system like, e.g., $^{132}$Sn (virtually) spends part of the time  in the ground state of $^{134}$Sn and part in that of $^{130}$Sn. This is another way to say that addition and removal pairing vibrations, are $J^\pi=0^+$, $\tau=1$, $\beta=\pm 2$ elementary modes of nuclear excitation, $\tau$ and $\beta$ being the isospin and the particle (fermion) quantum numbers. This last quantum number is associated with the number of particles operator $\hat N$, conjugate to the gauge \mbox{angle $\phi$,}
\begin{align}
[\hat N&,\phi]=i,\\
\hat N&=\sum_\nu a^\dagger_\nu a_\nu,
\end{align}
where $a^\dagger_\nu$ and $ a_\nu$ are the creation and annihilation operators of a nucleon moving in the state with quantum numbers labeled $\nu$ (e.g. $nljm$).
Another way to say the same thing is that pairs of nucleons moving in time reversal states $(\nu,\bar \nu)$ around (in) closed shells, eigenstates of the mean field, single--particle Hamiltonian
\begin{equation}
   H_{sp}=\sum_\nu (\varepsilon-\lambda) a^\dagger_\nu a_\nu,
\end{equation}
and correlating  through a pairing interaction
\begin{equation}
    H_p=-GP^\dagger P,
\end{equation}
where
\begin{equation}\label{eq9}
    P^\dagger=\sum_{\nu>0}a_\nu^\dagger a_{\bar\nu}^\dagger \quad \text{and} \quad P=\sum_{\nu>0}a_\nu a_{\bar\nu},
\end{equation}
are the pair addition and pair removal modes (pairing vibrations), give rise to (weakly correlated, spatially extended) two--particle (two--hole) like states, providing a microscopic description of the ground state of the systems with two more (less) nucleons than those in the closed shell system \cite{Bohr:64,Bes:66}. This is in keeping with the fact that the Hamiltonian
 \begin{equation}\label{eq2}
    H=H_{sp}+H_p,
\end{equation}
 is invariant under gauge transformations
\begin{equation}\label{eq6}
    \mathcal{G}(\phi)=e^{-i\frac{\hat N}{2}\phi},
\end{equation}
which is tantamount of saying that $[H,\hat N]=0$.

The two correlated neutrons moving on top of the core ($^{132}$Sn) Fermi surface as well as the two correlated neutron holes moving in this Fermi sea, can be viewed as nuclear embodiment of single Cooper pairs\cite{Cooper:56}.

\begin{figure}
\centerline{\includegraphics*[width=9cm,angle=0]{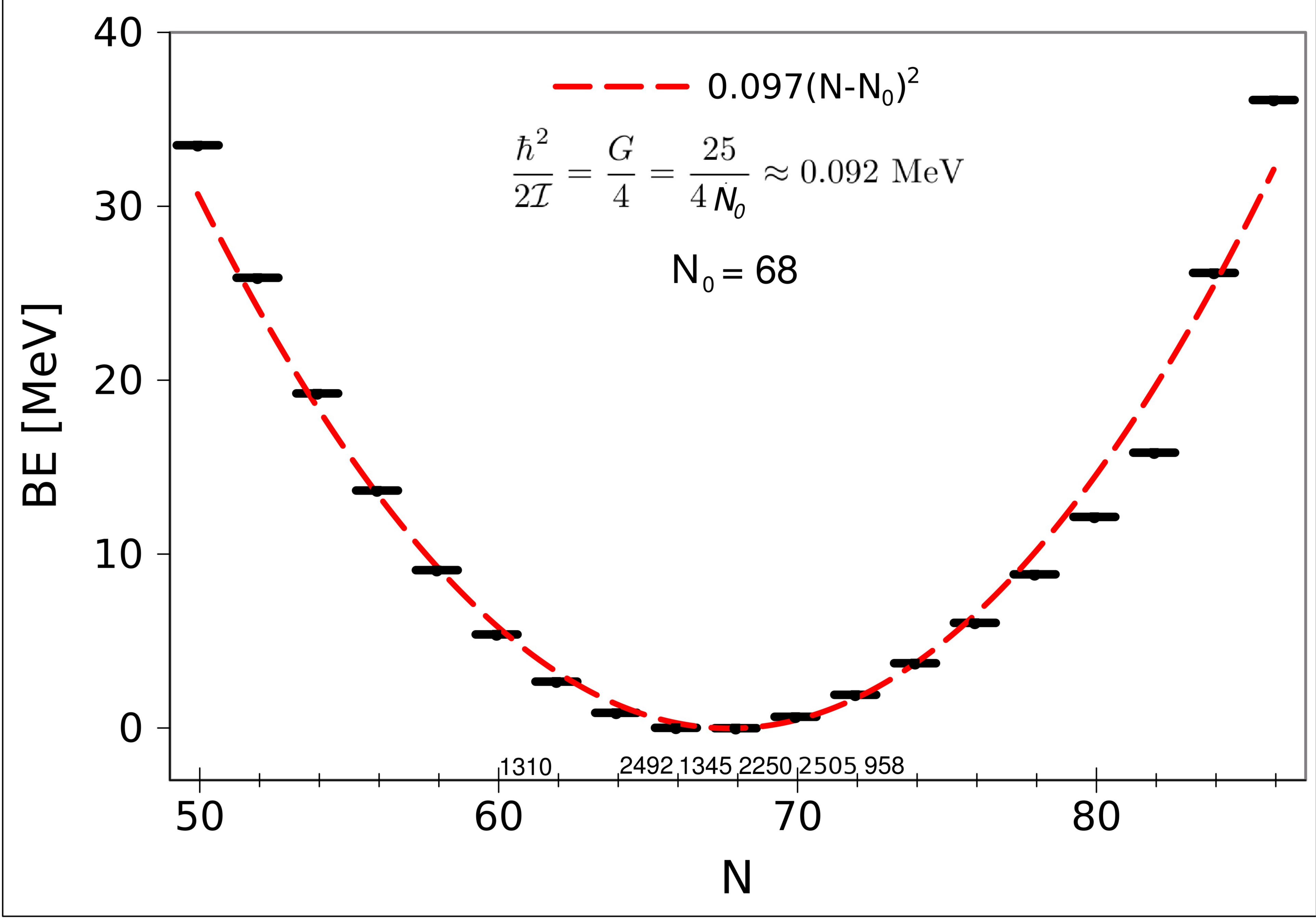}}
\caption{Ground state (pairing rotational spectrum) of Sn--isotopes. Also shown are the integrated ($p,t$) absolute cross  sections \cite{Guazzoni:99,Guazzoni:04,Guazzoni:06,Guazzoni:08,Guazzoni:11}, as well as predictions of the single $j$--shell model
}\label{fig1}
\end{figure}

\begin{figure}
\centerline{\includegraphics*[width=9cm,angle=0]{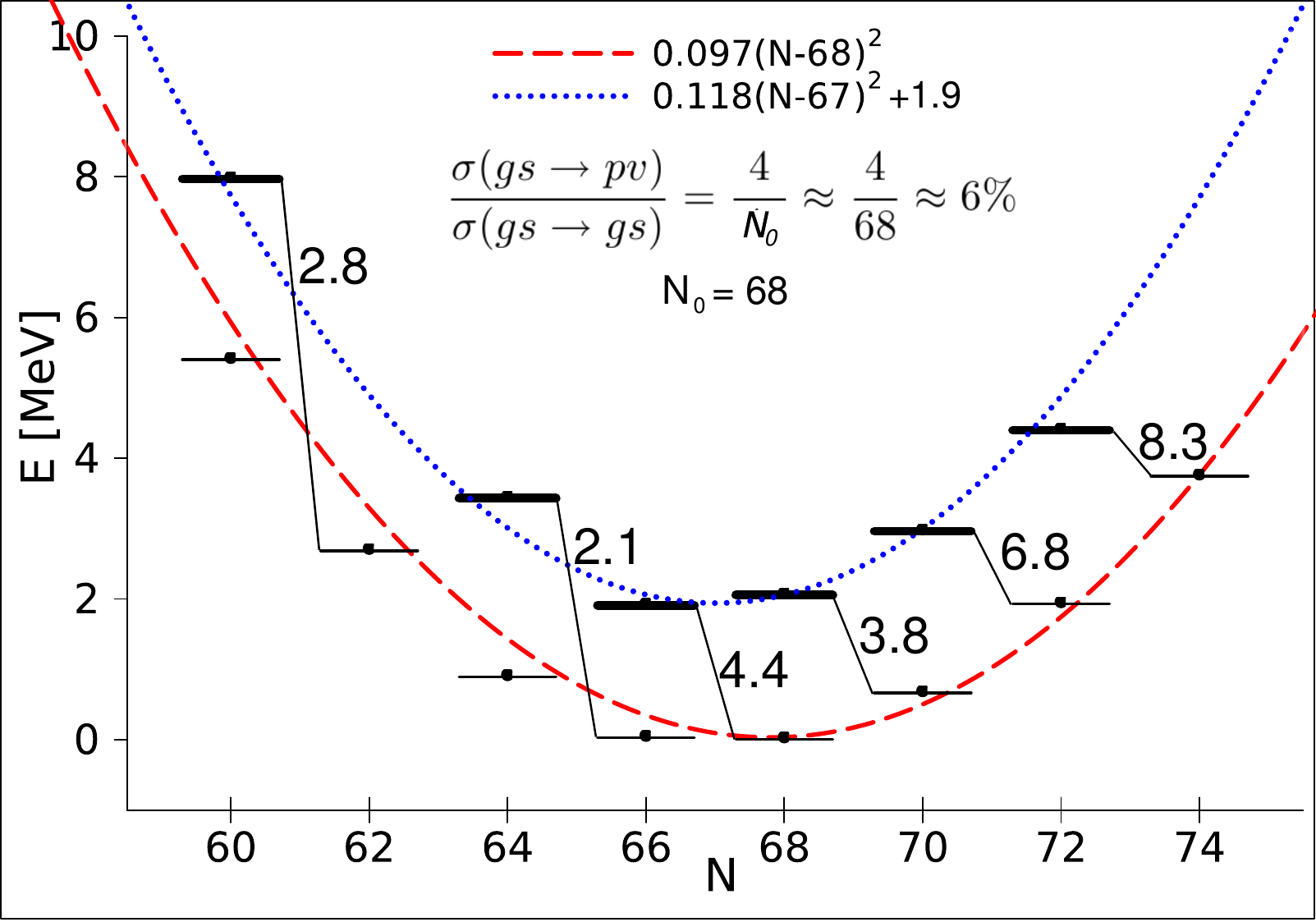}}
\caption{Ground state (dashed curve) and two--quasiparticle (pairing vibration; dotted curve)--based pairing rotational bands. The numbers given are the two--particle transfer cross sections populating the excited bands ($0^+$ with $E \lesssim 3$ MeV), normalized with respect to the gs$\rightarrow$gs transitions. The data is from refs \cite{Guazzoni:99,Guazzoni:04,Guazzoni:06,Guazzoni:08,Guazzoni:11}.
}\label{fig2}
\end{figure}

Moving away from closed shells towards the lighter Sn--isotopes will reduce the restoring constant associated with pairing vibrational modes and, eventually, in concomitance with its vanishing and associated Cooper pair condensation, the system will become deformed in gauge space, that is\cite{Bardeen:57a,Bardeen:57b},
\begin{align}\label{eq8}
_\mathcal{K}\langle BCS(\phi)|   P^\dagger|&BCS(\phi)\rangle_\mathcal{K}=\alpha_0=e^{i\phi}\alpha_0',\\
\alpha'_0&=\sum_{\nu>0}U_\nu V_\nu,
\end{align}
defining a privileged orientation in gauge space, and thus a gauge angle $\phi$,relating the intrinsic body--fixed reference system $\mathcal{K}'$, to the laboratory system $\mathcal{K}$, as testified by the BCS wavefunction
\begin{equation}\label{eq1}
    |BCS(\phi)\rangle_\mathcal{K}=|BCS(\phi=0)\rangle_\mathcal{K'}=\Pi_{\nu>0}(U_\nu+e^{-i\phi}V_\nu a_\nu^\dagger a_{\bar\nu}^\dagger)|0\rangle=\Pi_{\nu>0}(U_\nu+V_\nu a_\nu'^\dagger a_{\bar\nu}'^\dagger)|0\rangle,
\end{equation}
where
\begin{equation}\label{eq18}
a_{\nu}'^\dagger= \mathcal{G}(\phi)a_{\nu}^\dagger \mathcal{G}^{-1}(\phi)=e^{-i\frac{i}{2}\phi}a_{\nu}^\dagger.
\end{equation}
Relations (\ref{eq8})-(\ref{eq1}) testify to the phenomenon of spontaneous breaking of gauge symmetry, emergent properties\cite{Anderson:58} being frictionless flow, as testified by the small value of the moment of inertia of quadrupole deformed nuclei (see also the  Chapters contributed by Belyaev, Stephens and Lee and Frauendorf to this Volume), as well as generalized rigidity in gauge space (pairing rotational bands, see Figs. \ref{fig1} and \ref{fig2}; cf. Chapters contributed by Bes and by Hansen  to the present volume).
The state (\ref{eq1}) is the (ground state) solution, of the static, mean field approximation to the Hamiltonian (\ref{eq2}), that is 
\begin{equation}\label{eq3}
H_{MF}=H_{sp}+U_p,
\end{equation}
where
\begin{equation}\label{eq4}
U_p=-G\alpha_0(P^\dagger+P)-G\alpha_0^2.
\end{equation}
The state (\ref{eq1}) plays the role of the quasiparticle vacuum, that is,
\begin{equation}
\alpha_\nu'|BCS(\phi=0)\rangle_\mathcal{K'}=0,
\end{equation}
where
\begin{equation}\label{eq5}
\alpha_\nu'=U_\nu a_\nu'-V_\nu a_\nu'^\dagger,
\end{equation}
is the quasiparticle annihilation operator. There are two fields which correlate quasiparticles and which can eventually lead to time--dependent long range order (LRO), and thus to collective pairing modes. In the harmonic approximation these fields are: \textbf{a}) $(U_\nu^2-V_\nu^2)(\Gamma_\nu^\dagger+\Gamma_\nu  )$ leading essentially to a bound two--quasiparticle like state lying on top of the pairing gap, $\Gamma_\nu^\dagger=\alpha_\nu^\dagger\alpha_{\bar\nu}^\dagger$ being the two--quasiparticle (quasiboson) operator. It is of notice that in the case of closed shell nuclei (normal systems) this field gives rise to pair addition $(U_\nu^2=1,\;V_\nu^2=0; \varepsilon_\nu>\lambda)$, and to pair removal $(U_\nu^2=0,\;V_\nu^2=1; \varepsilon_\nu<\lambda)$ modes;
\textbf{b}) $(U_\nu^2+V_\nu^2)(\Gamma_\nu^\dagger-\Gamma_\nu  )$, which sets the $|BCS\rangle_\mathcal{K'}$ state into rotation in gauge space, and whose fluctuations diverge in the long wavelength limit (all different orientations of $|BCS\rangle_\mathcal{K'}$ have the same energy) in just such a way that the resulting ``exact'' ground state.
\begin{equation}\label{eq7}
    |N\rangle \sim \int d\phi\, e^{i\frac{N}{2}\phi}\,|BCS\rangle_\mathcal{K}\sim \left(\sum_{\nu>0}c(\nu)a_\nu^\dagger a_{\bar\nu}^\dagger\right)^\frac{N}{2}|0\rangle,
\end{equation}
transforms in an irreducible manner under the operator  (\ref{eq6}), having a fixed number of particles, gauge invariance being in this way restored.

 In other words, as the restoring force of the pair vibration vanishes in moving away from closed shells, the field \textbf{b}) leads to a long wavelength mode in which each cycle visits different orientations of the system in gauge space. And this is done with the right frequencies in $(\dot \phi=N/\hbar)$, as fixed by the value of the number of particles of the system. \textit{This is another example of the fact that while potential energy prefers spatial arrangements between particles, fluctuations, quantum or classical, prefer symmetries.}

The states $|N\rangle$ are the members of a nuclear pairing rotational band, Goldstone (\textbf{or better Anderson--Goldstone--Nambu (AGN)}) mode associated with the breaking of gauge invariance\cite{Anderson:58,Anderson:76,Nambu:60b,Nambu:60,Goldstone:61}. Within a schematic model (single $j$--shell), the associated energy can be written as (cf. e.g. app. H of ref \cite{Brink:05}),
 \begin{equation}\label{eq20}
    E=-\frac{G\Omega}{2}N+\frac{\hbar^2}{2\mathfrak{I}}N^2,
\end{equation}
where the moment of inertia is determined by the relation (see also Fig. \ref{fig1}),
 \begin{equation}
  \frac{\hbar^2}{2\mathfrak{I}}=\frac{G}{4}.
\end{equation}
At an energy,
 \begin{equation}
   2E=G\Omega,
\end{equation}
where $\Omega=(2j+1)/2$ is the pair degeneracy of the shell,  one finds the two--quasiparticle states. The ratio,
 \begin{equation}\label{eq19}
  \frac{\hbar^{2}/2\mathfrak{I}}{2E}\approx\frac{1}{\Omega},
\end{equation}
testifies to the fact that the pairing rotational excitation has an energy which is much smaller than that associated with the   intrinsic, quasiparticle energies. In keeping with the fact that, with some caveats (see discussion below on quadrupole pairing), pairing vibrations in superfluid nuclei can be viewed as bound two--quasiparticle states, lying on top of twice the pairing gap, the estimate (\ref{eq19}) can also be used in connection with these modes. In Figs. \ref{fig1} and \ref{fig2}  we display experimental evidence for the existence of pairing rotational bands, as well as for the validity of the relation (\ref{eq19}).
\section{Hindsight}
The ground state of a many--body quantum system in general, and of a finite many--body system (FMBS) like the atomic nucleus in particular, can be viewed as the vacuum state of a field theory, the Nuclear Field Theory (NFT) in the present case \cite{Bes:76a,Bes:76b,Bes:76c}. The fact that many--body Hamiltonians very often have ground states which do not have the same symmetry as the Hamiltonian itself is the broken symmetry phenomenon. In general, one breaks symmetry by introducing a field, which is taken not to have zero expectation value. This can be viewed as the order parameter, a physical variable, defined locally, which measures the distortion of the system under consideration. Such order parameters are usually not constants of motion, that is they do not commute with the Hamiltonian. Broken symmetry requires a Goldstone (AGN) boson, which in turn has zero point fluctuation which diverge in the long wavelength limit in such a way that symmetry is restored. In other words, in spite of the fact that the above mentioned singular behaviour leads to a divergent quasiparticle--phonon coupling, the theory can be renormalized resulting, among other things, in a finite moment of inertia of the pairing rotational band. These are some of the consequences of the fact that BCS is an asymptotic free theory. This   signifies two things: first, that something unexpected happens at the long wavelength limit and, second, that the high energy behaviour, although connected with logarithmic--like divergences, is harmless. Asymptotic free theories are not only renormalizable, they can be renormalized by the simple process of inventing a model, like the BCS model, which correctly describes the infrared behaviour of the complex system, introducing a physically sensible cutoff and coupling constant. Asymptotic freedom from this point of view is merely the statement that nothing above such a cutoff has any relevance to low energy physics (see \cite{Anderson:76}).

\section{Specific probe of BCS--like correlations: two--nucleon transfer}
The (ground state) pairing rotational bands is the Goldstone (AGN) mode associated  with the spontaneous breaking of gauge invariance in superfluid atomic nuclei, in a similar way in which rotational bands in quadrupole deformed nuclei are the Goldstone (AGN) modes associated with spontaneous breaking of rotational invariance\cite{Bohr:64,Bes:66,Bohr:75}, see also\cite{Broglia:73,Broglia:73b}. In both cases, the deformation defines a privileged orientation: in gauge space in the first case, in 3D--space in the second. The associated angles relating the body fixed frame of reference $\mathcal{K}'$ of the symmetry violating state, to the laboratory frame $\mathcal{K}$ being: the gauge angle $\phi$ in the first case (2D--rotation), the Euler angles $\Omega$ (3D--rotation) in the second case, as testified by the collective wavefunctions (\ref{eq8}) and\cite{Bohr:75}.

 \begin{equation}
 |IMK(\Omega)\rangle=\left(\frac{2I+1}{4\pi}\right)^{1/2}\mathcal{D}_{M K=0}^I(\Omega)|\text{Nilsson}(\Omega=0)\rangle_{\mathcal{K}'},
\end{equation}
respectively.\\
The above wavefunction is the product of the rotation matrix and of the Nilsson wavefunction\cite{Nilsson:55} describing the intrinsic state of the system (e.g. single--particles moving in an axially symmetric deformed Saxon--Woods potential, whose symmetry axis forms  an angle $\Omega$ with the laboratory system). In other words, we are confronted with the task of specifically probing the structure of atomic nuclei whose internal long--range order LRO \cite{Yang:62} parameter is a phase, the gauge phase $\phi$ in the case of $|BCS\rangle$, the Euler angles in the case of $|\text{Nilsson}\rangle$. Now, quantal fluctuations of the order parameters lead, in the absence of unsymmetrical external forces, to rotations in gauge ($\omega=\dot \phi= \lambda/\hbar$) and in 3D--space ($\omega=\dot \Omega= \hbar I/\mathcal{I}$) respectively, and thus to the restoration of the full original symmetry. The external forces (experiment) necessary to ``pin down'' these quantal fluctuations can only come from systems which themselves violate the given symmetry\cite{Anderson:64b}.

While one is accustomed to work with measuring instruments which themselves are not rotational invariant like e.g., a proton beam which, in the laboratory, defines a privileged orientation and can thus be used to set 3D--deformed nuclei into rotation, and thus measure e.g. the quadrupole deformation of the system, one does not usually have around devices displaying gauge space coherence. In other words, objects which are a wavepacket of states with different number of particles, with which one can set a superfluid 
nucleus into rotation in gauge space. It is of notice that the fingerprint of deformation of FMB system are rotational bands.

\begin{figure}
\centerline{\includegraphics*[width=15cm,angle=0]{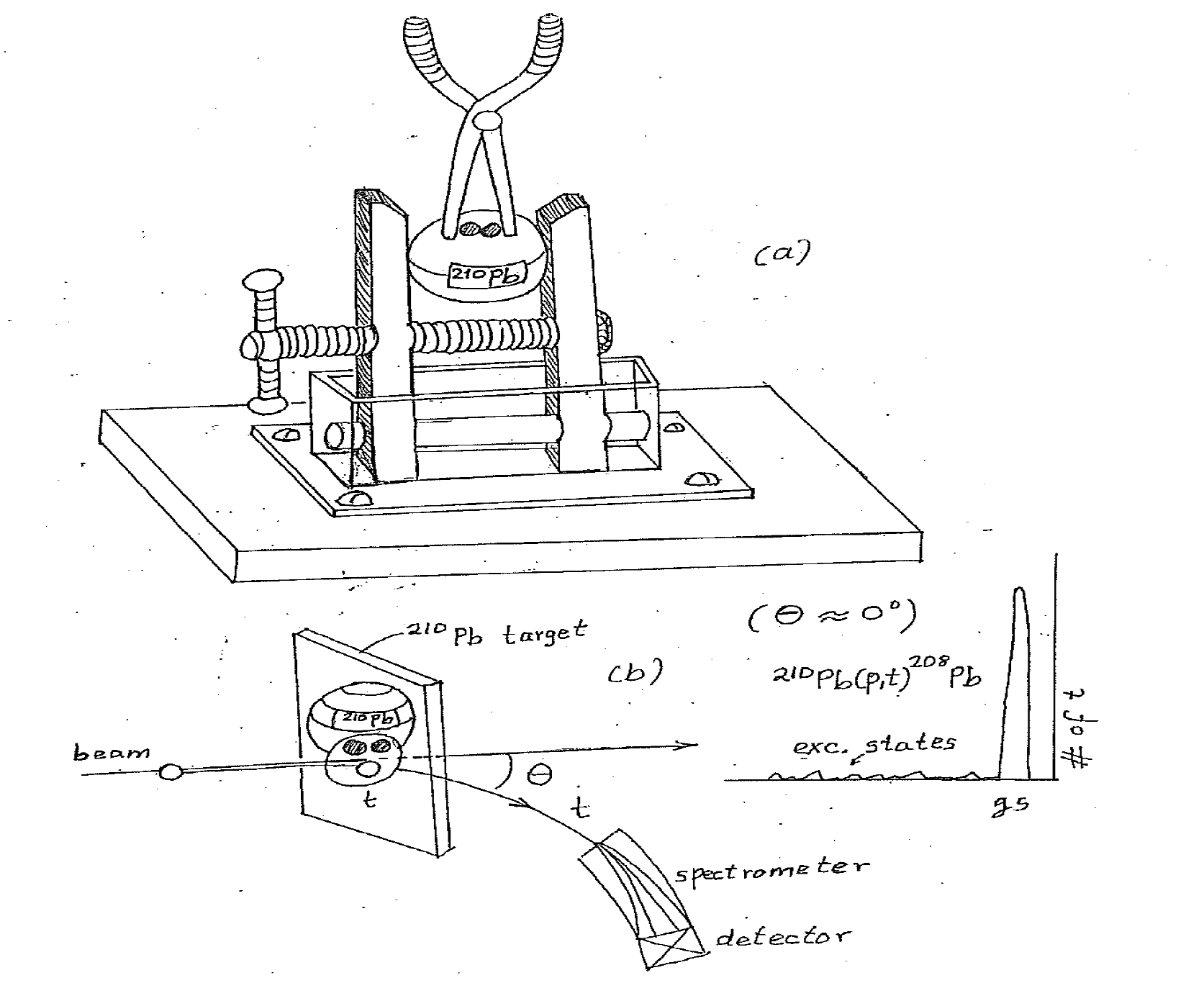}}
\caption{Cartoon illustrating that two--nucleon transfer reactions act as a ``clamp--tweezer'' device to pin down the (in this case, dynamical) gauge phase of Cooper pair pairing correlated nucleons, constituting the specific probe of pairing correlations in nuclei.}\label{fig3}
\end{figure}

The importance of the Josephson\cite{Josephson:62} effect, the superconducting tunnelling of electron Cooper pairs across a thin barrier (oxide layer) separating two superconductors, and leading to a DC current $J=J_1\sin(\phi_1-\phi_2)$ (AC current $J\sim\sin(\frac{2e}{\hbar}(\Delta Vt+2\delta))$ if biased), is that it provided for the first time an instrument, a clamp, which can pin down the (difference in) gauge phase existing between two superconducting systems. In fact, a metallic superconductor has a rather perfect internal gauge phase order (LRO), but the zero point motion of the total order parameter is large and rather rapid ($\dot \phi=N/\hbar$). Placing two such systems in weak coupling with each other, through Cooper pair transfer acting as tweezers, allows to pin down the (relative) gauge phase.
From the above narrative, one can posit that  two--particle, single Cooper pair, transfer processes provide the specific probe of  pairing correlation in atomic nuclei , and this not only to measure the LRO phase coherence of superfluid nuclei, but also the dynamic one, in connection with the excitation of pairing vibrational modes (see Fig. \ref{fig3}). In fact, although the $^1 S_0$ part of the correlation of the two neutrons in $^3_1 H_2$ is not identical to that correlating the neutron Cooper pair in e.g. $^{120}$Sn, a $(p,t)$ reaction on this target, is quite specific to probe of the associated BCS pairing correlations, as testified by the (schematic) expression of the corresponding absolute cross section connecting  members of the pairing rotational band  
 \begin{equation}\label{eq10}
\sigma \left(^{120}\text{Sn}(p,t)^{118}\text{Sn(gs)}\right)\sim |\alpha_0|^2=\left|\sum_{\nu>0}U_\nu(118)V_\nu(120)\right|^2,
\end{equation}
 proportional to the squared average value of the pair annihilation operator $P$ (see (\ref{eq9})). Within the single $j$--shell model of superfluid nuclei (see e.g. \cite{Brink:05} app. H),
 \begin{eqnarray}\label{eq11}
\sigma (\text{gs}\rightarrow\text{gs})&\approx \frac{A}{5},\\
\sigma (\text{gs}\rightarrow\text{pv})&\approx 1,
\end{eqnarray}
leading to the ratio
 \begin{eqnarray}\label{eq12}
R=\frac{\sigma (\text{gs}\rightarrow\text{gs})}{\sigma (\text{gs}\rightarrow\text{pv})}\approx\frac{A}{5},
\end{eqnarray}
which for Sn--isotopes ($A\approx120$)gives,
 \begin{eqnarray}\label{eq13}
R\approx 24.
\end{eqnarray}

The data displayed in Fig. \ref{fig2} provide evidence to the validity of the estimate (\ref{eq13}) and testify to the rather weak cross talk existing between pairing rotational bands based on different intrinsic states (ground state and pairing vibrational states below 3 MeV).

While supercurrents are not allowed in the essentially 0D--nuclear superfluid ($\xi/R\approx 30/6\approx 5$), the consequences of spontaneous symmetry breaking of gauge invariance in the FMB nuclear system can be studied in terms of individual quantal states, something which is not possible to do in connection with metallic superconductors.

In the first paper which addressed this subject\cite{Yoshida:62}, the relation (\ref{eq10})  was written as
 \begin{eqnarray}\label{eq14}
\sigma(gs \rightarrow gs) \sim (\Delta/ G)^2.
\end{eqnarray}
In keeping with the fact that within this simplified scenario, the cross section associated with the excitation of two--quasiparticle pairing vibrations is proportional to $V_\nu^2 (\varepsilon_\nu\approx\lambda)$, and thus $\sigma(gs \rightarrow pv)\approx 1$, it is   rather simple to estimate the expected enhancement factor (\ref{eq12}) ($\Delta\sim 1 $ MeV, $G\approx 0.2$ MeV, $R\sim 25$), which is in agreement with both (\ref{eq13}) as well as with the data (see Fig \ref{fig2}).

Nonetheless, the relation (\ref{eq14}), although in principle correct, is to be handled with care, so as not to incur in the misunderstandings  which were at the basis of the Bardeen--Josephson\cite{Bardeen:62} controversy in connection with what is now known as the Josephson effect, and which resulted in the eventual publication of Josephson's paper\cite{Josephson:62} as well as the paper of Cohen, Falicov and Phillips on pair tunnelling\cite{Cohen:62}. In other words, the emergence of Cooper pair transfer between superconductors as the specific probe of LRO gauge phase coherence. Also, of the development of the theoretical  and experimental tools to calculate and measure the associated (absolute) tunnelling supercurrents.

Within this context, it may prove illuminating to briefly recall the way in which, theoretical and experimental studies of pairing correlations in nuclei, have further shed light on the mechanisms which are at the basis of the structure and of the tunnelling of Cooper pairs. In particular, let us concentrate on: static (point \textbf{a}) below) and dynamic (point \textbf{b}) below) gauge phase LRO nuclear studies.

\textbf{a}) while supercurrents are not possible in atomic nuclei, in which the coherence length is larger than nuclear dimensions, it is possible to carry out single Cooper pair tunnelling experiments on superfluid, deformed nuclei in gauge space, thus increasing by one quantum at a time the rotational frequency in gauge space,  in term of individual quantal states (members of pairing rotational bands see  (\ref{eq7})	and (\ref{eq20}) and Figs \ref{fig1} and \ref{fig2});

\textbf{b})  pairing vibrations in metallic superconductors can be viewed as almost pure two--quasiparticle excitations. On the other hand, in normal nuclei, pairing vibrations, i.e. quantal fluctuations in particle number are quite collective. This is consistent with the fact of the  stronger role fluctuations play, as a rule, in FMBS as compared with (infinite) condensed systems (cf. ref. \cite{Broglia:73} and refs. therein).
   In keeping with the above narrative, the extension of the studies of tunnelling process of deformed systems in gauge space, to systems displaying \textit{dynamical} gauge LRO had to await for the measurement and calculation of two--nucleon transfer reaction cross sections in atomic nuclei (see the Chapter contributed by Hansen to this Volume). Within this context, i.e. collective vibrations associated with single Cooper pairs, it is of notice that one has found a new mechanism to correlate Cooper pairs, and eventually break gauge symmetry, which has emerged from studies of pairing correlations in exotic halo nuclei (cf. Sects. 6 and 7 below).

Let us come back for a moment to the question of the tunnelling between superconductors. Bardeen argued that virtual pair excitations in the superconducting ground state do not extend across the oxide layer\cite{Bardeen:62}. Thus, even if single electrons can tunnel through it, the tail of the pair wavefunction is virtually identical with that of the normal state.
Now, the crucial point in Cooper tunnelling, however, is that vanishing of the gap $\Delta$ does not imply the vanishing of $\alpha_0(\mathbf{r}, \mathbf{r}')$. On the contrary, $\alpha_0(\mathbf{r}, \mathbf{r}')$ can have large amplitude for fermions separated by distances $|\mathbf{r}- \mathbf{r}'|$ up to the coherence length $\xi=\hbar v_F/2\Delta$ (see e.g.\cite{Gorkov:59}).

Within this context one can mention the case of quadrupole pairing and of gapless superconductivity in, quadrupole deformed, actinide nuclei\cite{Casten:72,Ragnarsson:76,Vanrij:72}. In these nuclei, one observes a dynamic decoupling between the condensate and particular Nilsson orbitals lying close to the Fermi energy and displaying a value of the intrinsic quadrupole moment of sign opposite from that of the majority of the single--particle levels lying around the Fermi energy. In keeping with the fact that quadrupole deformed nuclei break rotational invariance, the ``effective'' pairing coupling constant, is the sum of the monopole and of the quadrupole pairing. Thus the interaction correlating Cooper pairs moving in oblate $(o)$--prolate $(p)$ Nilsson orbits

 \begin{equation}
G_{op} = G_0 - G_2|q_{\nu_o}q_{\nu_p}|,
\end{equation}
is much smaller than that which pairs particles moving in Nilsson states displaying the same sign of the intrinsic quadrupole moment. In other words
 \begin{equation}
G_{oo}\approx G_{pp}=G_0+G_2|q_{\nu}|^2\gg G_{op}, \quad (\nu=p,o).
\end{equation}
Consequently, pairs of nucleons moving in ($\nu_o,\nu_p$)--orbitals feel an essentially zero pairing gap. They thus lead  to pairing vibrations displaying cross sections relative to the ground state which are much larger than the expected value $R^{-1}\approx 5/A$ of typical superfluid nuclei (see (\ref{eq12})).

Let us now concentrate on the value of the amplitude for two fermions, at $\mathbf{r}$ and $\mathbf{r}'$ to belong to a Cooper pair namely $\alpha'_0(\mathbf{r},\mathbf{r}')$, that is, the nuclear structure component of the two--particle transfer cross section amplitudes (Cooper pair wavefunction). To clarify the physics which is at the basis of the renormalizability of the BCS description of the spontaneous gauge symmetry breaking of superfluid nuclei, in terms of a coupling constant $G$ and of an energy cutoff, we discuss two scenarios for the case of the $^{120}$Sn$(p,t)^{118}$Sn reaction. In the first one, we consider all bound single--particle states, the cutoff energy being $E_{cutoff}=0$ MeV. In the second case one sets $E_{cutoff}=60$ MeV, discretizing the continuum inside a spherical box of 8 fm of radius. The BCS gap ($\Delta=1.41$ MeV; experimental value) and number ($N=70$) equations lead to $G=0.51$ MeV and $\lambda=-\,6.72$ MeV in the first case and $G=0.22$ MeV and $\lambda=-\,6.9$ MeV in the second one. 
The associated Cooper pair probability distributions in r-space are essentially identical. It is then not surprising that they lead to very similar absolute values of the two--particle transfer cross section associated with the reaction $^{120}$Sn$(p,t)^{118}$Sn(gs).

 Let us now repeat the argument, but this time in terms of $(\Delta/G)^2$ (see Eq. (\ref{eq14})). Because the pairing gap has been fixed to reproduce the experimental value (1.41 MeV), one obtains in the case of $E_{cutoff}=0$MeV $R\sim 7.6$ and $R\sim 45$ in the case of  $E_{cutoff}=60$MeV. This result emphasizes the problem of working with an expression which contains explicitly the pairing coupling constant as (\ref{eq14}).

 One could argue that such an objection could also be leveled against the relation $\sigma\sim|\alpha_0|^2$. Note however, that a $(p,t)$ reactions would hardly feel the effect of contributions far removed from the Fermi surface $\lambda$. This is in keeping with the fact that transfer to levels lying far away from $\lambda$  will be unfavourable due to $Q$--value effects. If one argues in terms of the relative distance $r$ between target and projectile ($r\gg R_0$ for continuum--like contributions; $r<R_0$ for deeply bound--like contributions), the outcome is similar. In fact, for large distances the two--particle transfer formfactor vanishes while, at small distances the outgoing tritium will experience strong absorption.

In fact, considering only the contribution to $\alpha'_0$ arising from the valence orbitals, that is, essentially those contributing to the ``naked'' vision of the Cooper pair wavefunctions, one obtains $\alpha'_0=2.12$ and $\alpha'_0=2.08$ respectively, and thus, a negligible squared relative difference between the two predicted cross sections, namely $(0.04/2.1)^2\approx 2\times 10^{-3}$.

\section{Hindsight}
One can posit that while there exist accurate relations to implement renormalization protocols, arguably, they can  hardly take the place of physical intuition in renormalizing asymptotically free theories, like for example nuclear BCS (choice of adequate values of $G$ and of $E_{cutoff}$). Second, that the resulting (gauge invariant) coherent states (pairing rotational band) are extremely simple macroscopic--like quantal objects (almost classical states), whose wavefunctions can be calculated quite accurately and almost effortlessly, making use of schematic Hamiltonians (like e.g. the mean field BCS Hamiltonian). Within this context, the corresponding two--nucleon transfer spectroscopic amplitudes can be considered essentially accurate. Consequently, they can be used, together with the experimentally determined optical potentials, to quantitatively test the validity of two--nucleon transfer reaction theories in predicting absolute differential cross sections. This is also true concerning QRPA states associated with the dynamical violation of gauge invariance (pairing vibrations\cite{Bes:66}).

\section{Meeting theory with experiment: absolute two--particle transfer cross sections}
Although not experimentally easy to handle in trying to extract information concerning e.g. the absolute cross section associated with members of the (Sn--isotope) pairing rotational band, the reaction
\begin{equation}\label{eq16}
^{A}\text{Sn}+^{A'+2}\text{Sn}\rightarrow ^{A+2}\text{Sn}(\text{gs})+^{A'}\text{Sn}(\text{gs}),
\end{equation}
for values of $A$ away from 100 and 132, is the specific probe of pairing correlations in Sn--isotopes (think of the Josephson junction; see the contribution of von Oertzen to this Volume).
In this heavy ion collision, transfer reactions will take place together with strong Coulomb excitation processes. The analysis of such reactions can, as a rule, only be made in terms of a coupled--channel treatment .

Even if Coulomb excitation is weak it is not obvious that lowest order perturbation theory is adequate. For bombarding energies above the Coulomb barrier there will necessarily occur collisions with impact parameters which lead to strong interactions where the transfer process can no longer be treated to lowest order. The probability for populating a definite two--particle transfer channel, like e.g. the one written in (\ref{eq16}), can be written as\cite{Broglia:04a}

\begin{equation}
P_i=p_2(i)\Pi_{n\neq i}(1-p_n)\approx p_2(i)P_0,
\end{equation}
where
\begin{equation}
p_2(i)=|a_2(i)|^2,
\end{equation}
$a$ standing for a transfer amplitude calculated to second order of perturbation theory. The damping factor
\begin{equation}
P_0=\exp\left(-\sum_n p_n\right),
\end{equation}
can be included by using an imaginary part in the (relative motion) potential (optical potential $U+iW$). It is of notice that one can also calculate microscopically $W$.

The channels explicitly considered in the calculation of $a_2$ are: $\alpha \equiv (A+a(=b+2))\rightarrow (B(=A+2)+b)\equiv \beta$, the intermediate channel $\gamma$ corresponding to  one--nucleon transfer channel $F(=A+1)+f(=b+1)$. The associated amplitudes are, in the semiclassical approximation (prior--prior representation)
\begin{equation}
(a_\beta)_{(1)}=\frac{1}{i\hbar}\int_{-\infty}^{\infty}dt \langle \Psi_\beta|V_\alpha-\langle V_\alpha\rangle|\Psi_\alpha\rangle_{\mathbf{R}_{\beta \alpha}}e^{i(E_\beta-E_\alpha)t/\hbar},
\end{equation}
\begin{multline}
(a_\beta)_{succ}=\left(\frac{1}{i\hbar}\right)^2\int_{-\infty}^{\infty}dt \langle \Psi_\beta|V_\gamma-\langle V_\gamma\rangle|\Psi_\gamma\rangle_{\mathbf{R}_{\beta \gamma}}e^{i(E_\beta-E_\gamma)t/\hbar}\\
 \times\int_{-\infty}^{t} dt' \langle \Psi_\gamma|V_\alpha - \langle V_\alpha \rangle|\Psi_\alpha\rangle_{\mathbf{R}_{\gamma \alpha}}e^{i(E_\gamma-E_\alpha)t'/\hbar},	
\end{multline}
and
\begin{multline}
(a_\beta)_{orth}=-\frac{1}{i\hbar}\sum_{\gamma\neq(\beta,\alpha)}\int_{-\infty}^{\infty}dt \langle \Psi_\beta|\Psi_\gamma\rangle_{\mathbf{R}_{\beta \gamma}}
  \langle \Psi_\gamma|V_\alpha -\langle V_\alpha \rangle|\Psi_\alpha\rangle_{\mathbf{R}_{\alpha\gamma }}e^{i(E_\beta-E_\alpha)t/\hbar}.
\end{multline}
The first term $(a_\beta)_{(1)}$, describes the simultaneous transfer of two nucleons. The two--step successive transfer is described by $(a_{\beta})_{succ}$, while $(a_{\beta})_{orth}$ is a second--order contribution arising from the non--orthogonality of the channels considered.

In the independent particle limit this term cancels exactly  the simultaneous transfer contribution $(a_\beta)_{(1)}$. The transfer reaction is then described as a purely successive process by the amplitude $(a_{\beta})_{succ}$, as was expected. Under normal, physical circumstances, much of this cancellation will still be operative, in keeping with the fact that the correlation energy binding Cooper pairs is weak ($\sim 1$ MeV) as confronted with the Fermi energy $\varepsilon_F(\approx 36$MeV).

  Another wording of the same concept is based on the fact that pairing can be viewed as a small correction to the independent particle motion $(G/|U(r=R_0)|\approx 0.2$ MeV/25 MeV $\approx 10^{-2}, U$ being the single--particle potential, while $G$ is the pairing coupling constant, $G\approx 25$MeV $/A$). Further circumstantial evidence to this cancellation is provided by the fact that extending the summation appearing in $(a_{\beta})_{orth}$ to all states of the two nuclei $f$ and $F$ (including continuum states), one can then perform closure in which case $(a_{\beta})_{orth}$ exactly cancels out $(a_\beta)_{(1)}$.


The other limiting situation which can be analyzed in simple terms, namely that in which the two transferred particles are strongly correlated, can be better understood rewriting the above amplitudes in the so--called mixed, post--prior representation, just an embodiment of energy conservation (i.e, $H=T_{aA}+H_a+H_A+V_{aA}(\equiv T_\alpha+H_\alpha+V_\alpha)=T_\gamma+H_\gamma+V_\gamma=T_\beta+H_\beta+V_\beta)$. In this representation, the non--orthogonality term is zero (and of course the new expression of $(a_\beta)_{(1)}\sim \langle \Psi_\beta|V_\beta-\langle V_\beta\rangle|\Psi_\alpha\rangle_{\mathbf{R}_{\beta \alpha}}$ smaller than the previous one).

In the case in which $V_{12}=G\rightarrow \infty$, a very large energy will be associated with the breaking of a pair in the intermediate $\gamma(=f+F))$ channel (i.e. $(E_\beta-E_\gamma),(E_\gamma-E_\alpha)\gg (E_\beta-E_\alpha)$), the corresponding exponentials leading to rapid oscillations and thus to a cancellation of the two orbital integrals appearing in $(a_{\beta})_{succ}$. In this case, the transfer of two particles thus takes mainly place as a simultaneous transfer. In nuclei, the first scenario (independent particle scenario), is closer to the actual experimental situation than this last one.

It is of notice that there is an intimate connection between the expression of $(a_\beta)_{(2)}$, and that of the matrix elements of the tunnelling Hamiltonian

\begin{equation}\label{eq17}
H_T=\sum_{kq\sigma}\left(T_{kq}a^\dagger_{k\sigma}a_{q\sigma}+T_{qk}a^\dagger_{q\sigma}a_{k\sigma}\right),
\end{equation}
used in connection with the Josephson  effect (cf. \cite{Anderson:64b,Cohen:62}). In the above expression $T_{kq}$ is an exponentially small tunnelling matrix element from state $k$ on one side of the oxide layer to the state $q$ on the other side, that is $a(=f+1)$ and $A(=F+1)$ within the nuclear context.

The $U_k V_q$, BCS coherence factor associated with two--nucleon transfers reactions arises here in connection with the fact that
\begin{eqnarray}
&a_k^\dagger(A) a_q(a) |BCS(A), BCS(a)\rangle\\
&=U_k V_q \alpha_k^\dagger(A)\alpha^\dagger_q(a) |BCS(A),BCS(a)\rangle
=U_k V_q |1qp_k(F),1qp_q(f)\rangle.
\end{eqnarray}
Within this context, Bardeen\cite{Bardeen:62} was right concerning the fact that the process described by the amplitude $(a_\beta)_{(1)}$ could hardly give rise to observable effects. On the other hand, because the coherence length of Cooper pairs in metals is larger than the oxide layer thickness, the order parameter $\sum U_kV_q$ can be built up, without loss of coherence, one step at a time, with the help of the above expression, namely transference of one fermion at a time. A dynamical embodiment of this mechanism is, arguably, at the basis of the bootstrap binding of the halo Cooper pair to the core of $^9$Li, leading to the weakly bound $^{11}$Li nucleus (see below).
\begin{figure}
\centerline{\includegraphics*[width=12cm,angle=0]{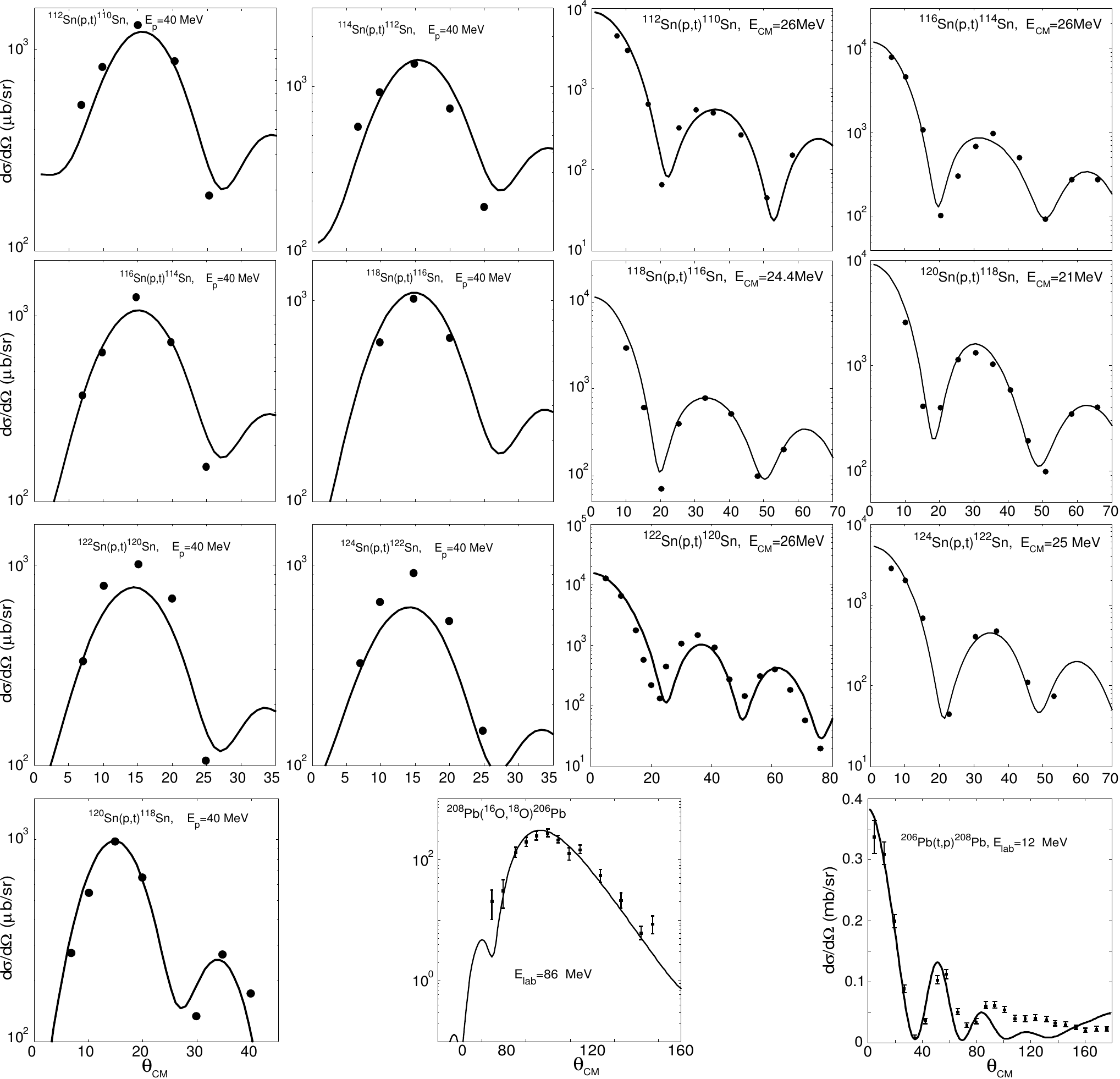}}
	\caption{Absolute values of two--particle differential cross sections calculated as explained in the text, in comparison with experimental results \cite{Guazzoni:04,Guazzoni:06,Guazzoni:08,Guazzoni:11,Guazzoni:99,Bassani:65,Bayman:82}, Bjerregaard et al. Nucl. Phys. \textbf{89},337,(1966).}\label{fig8}
\end{figure}

Because of spatial quantization, it may be possible, through strongly negative $Q$--value effects, to make very unfavourable one--particle transfer, and thus the successive transfer of two nucleons, like e.g. in the isotopes of Sn with mass number $A=132$, and of Be ($A=12$). By properly adjusting the bombarding energy, one would provide optimal conditions for simultaneous transfer. Whether the associated weak absolute cross section can be recorded at profit, is an open question.

Let us remind that the overcompleteness and thus non--orthogonality displayed by the elementary modes of excitation basis employed in NFT to describe nuclear structure leads to a coupling Hamiltonian to be diagonalized, as a rule, perturbatively, in keeping with the fact that most of the nuclear correlations are already included in the basis states (i.e. single--particle and collective motion). Similarly, the tunnelling Hamiltonian (\ref{eq17}) can be derived by finding sets of single--particle functions for each side of the oxide junction (left and right superconductors) in the absence of the potential of the other metal. Then, one eliminates nonorthogonality effects by perturbation theory\cite{Anderson:64b}.

While many analogies can be established between condensed matter BCS related phenomenon and FMBS nuclear ones, there are also profound differences. In particular those emerging from spatial quantization, leading to a discreteness of the spectrum which allows to study the effects of increasing the rotational frequency in gauge space (particle number) in terms of individual levels. For this purpose, one has to recur to light ion induced reactions and thus to quantal, DWBA calculation of the successive, simultaneous and  non--orthogonality contributions. Within the above scenario a quite homogeneous, state of the art set of $(p,t)$ data $(E_{CM}\approx25$ MeV) along the Sn--isotopes exists\cite{Guazzoni:04,Guazzoni:06,Guazzoni:08,Guazzoni:11,Guazzoni:99}. In Fig. \ref{fig8} the absolute differential cross section associated with transitions along the ground state pairing rotational band are shown in comparison with the calculations\cite{Potel:11} carried out making use of the BCS--based two--nucleon transfer spectroscopic amplitudes, and of the optical potential used in the experimental work \cite{Guazzoni:06}, plus those of ref \cite{An:06} for the $f+F$ (one--nucleon transfer, deuteron) channel. Theory agrees with data within the experimental errors or better. It is of notice that when the two-nucleon transfer spectroscopic amplitudes are calculated with the help of an extended shell model wavefunction obtained making use of realistic matrix elements (cf. e.g. \cite{Guazzoni:04} as well as the contribution of Covello, Gargano and Kuo to the present volume), one obtains  results which can hardly be distinguished from the BCS--based predictions. This is another example of the simplicity displayed by coherent states at large, and in gauge space in particular, states which behave almost semiclassically.

 Making use of the same input with regard to both two--nucleon spectroscopic amplitudes and optical potentials, but changing the proton bombarding energy to 40 MeV, one has recalculated again the $^{A+2}$Sn$(p,t)^A$Sn(gs) absolute differential cross sections. The corresponding results in comparison with the experimental data\cite{Bassani:65}, are also displayed in Fig.\ref{fig8}. Again, in this case, theory accounts for the experimental findings within errors (see also Table 1 of the contribution of Broglia to this Volume). Also shown in Fig. \ref{fig8} are the absolute differential cross sections associated with the $^{206}$Pb $(t,p)$ and $^{208}$Pb$(^{16}$O,$^{18}$O) excitation of the pair removal mode of $^{208}$Pb, calculated making use of RPA wavefunctions (two--nucleon spectroscopic amplitudes, cf. \cite{Broglia:73} and references therein), and of global optical parameters . Theory again provides a quantitative account of the data.
From the results displayed in Fig. \ref{fig8}, it seems fair to posit that two--nucleon transfer reaction theory has reached a quantitative level.

Within this context, it is important to remind of the fact that many groups have contributed through the years to develop the reaction theory of two--nucleon transfer processes including simultaneous, successive and non--orthogonality contributions into a tool to calculate the absolute differential cross sections which can be directly compared with the experiment findings (see \cite{Udagawa:73,Chien:75,Segawa:75,Schneider:76,Gotz:75,Takacsy:74,Hashimoto:78,Kubo:78,Bayman:82,Yagi:79,Igarashi:91,Becha:97} and refs. therein; see also the Chapter of Thompson in this Volume).

\subsection{Hindsight}
BCS theory, arguably like QED, belongs to a class of descriptions of physical phenomena which come close to certainty. This is not so much because they can be microscopically derived from the ground up without free parameters or divergences -think only of $G$ and $E_{cutoff}$ in the first case and of renormalization in the second- but primarily because of the wide variety of phenomena they can correlate,  the Josephson effect and the Lamb shift providing textbook examples. Not only this, but also the fact that they contribute paradigms which carry over other fields of research not thought of in the first time.

Because the BCS spectroscopic amplitudes describing the two--nucleon transfer process along a pairing rotational band associated with the valence orbitals can be considered essentially ``exact'', together with the fact that global studies of elastic scattering have led to reliable optical model parameters for the different channels involved, it is possible to test the nuclear tunnelling reaction mechanism quite accurately. As testified by the fact that theory provides, within experimental errors, an overall account of the absolute differential cross sections for a rather large sample of the available transfer data, one can posit that the 2nd--order DWBA two--nucleon transfer reaction mechanism  including successive, simultaneous and non--orthogonality contributions, provides a quantitative description of single Cooper pair nuclear tunnelling.


\section{Searching for the sources of BCS condensation in nuclei: measuring phonon induced pairing with single Cooper pair transfer}

The $N=6$ isotope of $_3^9$Li displays quite ordinary structural properties and can, at first glance, be thought of a two--neutron hole system in the $N=8$ closed shell. That this is not the case emerges clearly from the fact that $^{10}$Li is not bound, the lowest virtual ($1/2^+$) and resonant ($1/2^-$) states testify to the fact that, in the present case, the $N=6$ is a far better magic neutron number in the present case than $N=8$. Furthermore, that the unbound $s_{1/2}$ state lies lower than the unbound $p_{1/2}$ state, in plain contradiction with \textit{static} mean field theory.

\begin{figure}
\centerline{\includegraphics*[width=15cm,angle=0]{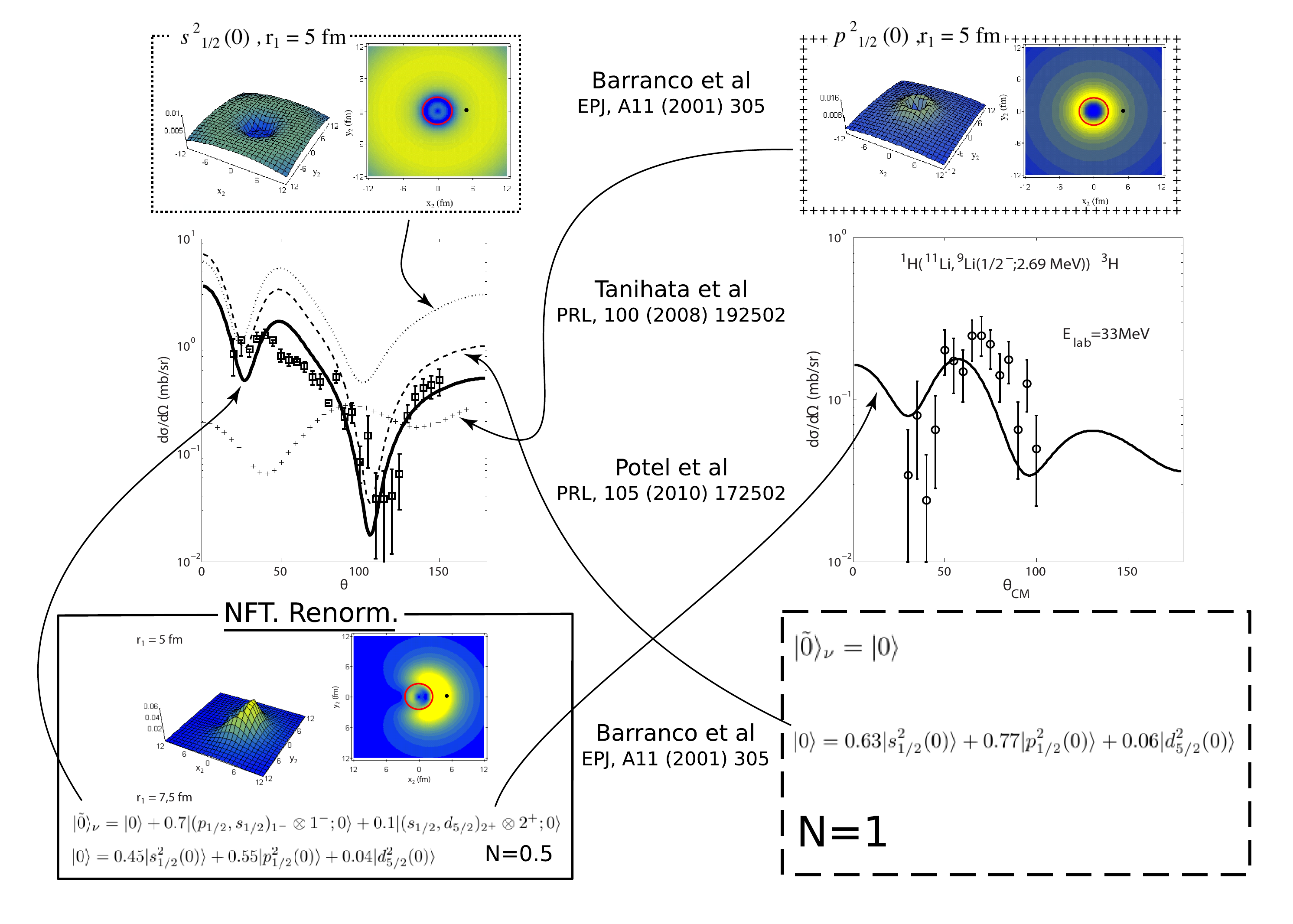}}
\caption{Absolute, two--nucleon transfer differential cross section associated with the ground state and the first excited state of $^9$Li, excited\cite{Tanihata:08} in the reaction $^1$H($^{11}$Li,$^9$Li)$^3$H in comparison with the predicted differential cross sections\cite{Potel:10} calculated making use of spectroscopic amplitudes and Cooper pair wavefunctions calculated in NFT.}\label{fig11}
\end{figure}
Dressing the (standard) mean field single--particle state with vibrations (dynamic shell model, see e.g. \cite{Mahaux:85} and refs. therein), mostly with the core quadrupole vibration, through polarization (effective mass--like) and correlation diagrams (vacuum zero point fluctuations (ZPF)) diagrams, similar to those associated with the (lowest order) Lamb shift Feynman diagrams), move the ${s}_{1/2}$ and ${p}_{1/2}$ mean field levels around. In particular the $1p_{1/2}$ from a bound state ($\approx -1.2$MeV) to a resonant state lying at $\approx 0.5$MeV (Pauli principle, vacuum ZPF process), and the $s_{1/2}$ continuum state down to an essentially bound, virtually stable (self--energy like diagrams), as  shown in ref.\cite{Barranco:01} (see also the contribution of Broglia to this volume).

Adding one further nucleon leads to a bound system. In fact, $^{11}_3$Li$_8$ displays a two--neutron separation energy $S_{2n}\approx 400$keV. A NFT description of this system which provides a quantitative, overall account of the experimental findings,(see \cite{Barranco:01} and references therein) testifies to the fact that the glue binding the neutron Cooper pair to the $^9$Li closed shell system are the core quadrupole vibration, and the dipole pigmy resonance resulting from the sloshing back and forth of the neutron halo with respect to the core protons, the bare $NN$--interaction playing a small role in determining the neutron Cooper pair structure, as testified by the wavefunction\cite{Potel:10}
\begin{equation}\label{eq21}
|0\rangle_\nu=|0\rangle+\alpha|(p_{1/2},s_{1/2})_{1^-}\otimes 1^-;0\rangle+\beta|(s_{1/2},d_{5/2})_{2^+}\otimes 2^+;0\rangle,
\end{equation}
with
\begin{equation}\label{eq22}
\alpha=0.7,\quad \text{and} \quad \beta=0.1,
\end{equation}
and
\begin{equation}\label{eq23}
|0\rangle=0.45|s_{1/2}^2(0)\rangle+0.55|p_{1/2}^2(0)\rangle+0.04|d_{5/2}^2(0)\rangle,
\end{equation}
the states $|1^-\rangle$ and $|2^+\rangle$ being the (RPA) states describing the dipole pigmy resonance of $^{11}$Li and the quadrupole vibration of the core. While these states are virtual excitations which, exchanged between the two neutrons bind them to the Fermi surface provided by the $^9$Li core, they can be forced to become real with the help of the specific probe of Cooper pairs in nuclei, namely two--particle transfer reactions.

Within this context, it is revealing that, the two final states populated in the inverse kinematics, two--neutron pick up reaction $^1$H($^{11}$Li,$^9$Li)$^3$H are, the $|3/2^-\text{gs}(^9\text{Li})\rangle$ and the first excited $|1/2^-,2.69\text{MeV}\rangle$\cite{Tanihata:08}. The associated absolute differential cross sections thus probe, within the NFT scenario, the $|0\rangle$  and the $|(s_{1/2},d_{5/2})_{2^+}\otimes 2^+;0\rangle$ component of the Cooper pair wavefunction respectively, the $p_{3/2}$ proton acting as a spectator. It is of notice that the $|1/2^-,2.69\text{MeV}\rangle$ state of $^9$Li can be viewed as the $1/2^-$ member of the multiplet resulting from the coupling of the $^8$He core quadrupole vibration and the $p_{3/2}$ proton.
Theory is compared with the experimental findings in Fig. \ref{fig11}. It reproduces the absolute two--particle differential cross section within experimental errors.

While no theory, let alone NFT is able to predict a single small amplitude of a wavefunction like $\beta$ with great accuracy (due essentially to the limited experimental information concerning the corresponding collective state), it can with uniqueness signal whether a rare channel is open or not. Because detailed, second order calculations of inelastic, break up and final state interaction channels, which in principle can provide alternative routes to the  $|1/2^-,2.69\text{MeV}\rangle$ state than that predicted by the NFT  ($\beta$ component), lead to absolute cross sections which are smaller by few orders of magnitude than that shown in Fig. \ref{fig11} (excited state)\cite{Potel:10}, one can posit that quadrupole core polarization effects in $|\text{gs}(^{11}\text{Li})\rangle$, are essential to account for the observation of the $|1/2^-,2.69\text{MeV}\rangle$ state.
\subsection{Hindsight}
Essentially three decades ago, the observation of the $^{14}$C decay of $^{223}$Ra, leaving behind the almost doubly magic nucleus $^{209}$Pb was reported in the literature\cite{Rose:84}. This observation started a flurry of activity to individuate and explain exotic decay, as the phenomenon was called (see e.g.\cite{Price:89}). The measured decay constant $\lambda=4.3\times10^{-16}s^{-1}$ testifies to the fact that the wavefunction describing the ground state of $^{223}$Ra had a component which could be viewed as $^{14}$C and a $^{209}$Pb nucleus essentially in contact, with a probability of the order of $P_{form}=10^{-10}$. It was shown that if one assumes normal nuclear matter to fill the nucleus $^{223}$Ra, $P_{form}$ differed from that required by experiment by many orders of magnitude, while the empirically determined number emerged naturally taking into account the fact that the emitor is superfluid. Of course, the calculations could hardly do better than to predict $P_{form}$ within one or two orders of magnitude accuracy (see Chapter contributed by Bertsch to the present Volume as well as Ch. 7 of \cite{Brink:05} and refs. therein). Nonetheless one can posit with reasonable degree of certainty that superfluidity was what is needed to explain the new channel through which nuclei can transmute one species into another one through the emission of a light nucleus. Within this context is that one can assess the role of the small amplitude $\beta$ (see eq. (\ref{eq22})) with which the quadrupole vibration of the $^8$He core is present in the ground state wavefunction of $^{11}$Li, rather than through its actual, obviously uncertain, value. It is of notice that this is a novel embodiment  (bootstrap phonon mediated pairing) of the Bardeen--Fr\"{o}hlich--Pines mechanism to produce Cooper condensation and thus break gauge invariance.

\section{Search of novel pairing modes}
The pairing vibration spectrum has been studied in detail in few regions of the mass table, in particular around the doubly magic nucleus $^{208}$Pb, where states up to three pairing vibrational phonons were observed, albeit displaying non negligible amounts of anharmonic effects (see \cite{Flynn:72} and references therein).

The availability of beams of the doubly magic nucleus $^{132}$Sn opens new possibilities to shed light on the mode--mode interactions in connection with pairing vibrational modes around this newly discovered doubly magic nucleus\cite{Jones:10}. Another extremely interesting double closed shell system is $^{100}$Sn, although it is an open question how close one can come to such a system and carry out measurements on it. As already testified by $N=Z$ doubly closed shell systems like $^{16}$O and $^{40}$Ca, coexistence, that is the presence of deformed excited ($4p-4h,2p-2h$) $0^+$ states, in spherical nuclei, allows to probe the system, also through two--nuclear transfer reactions, in particular concerning the interplay of pairing and deformation and, eventually also (quadrupole) phonon mediated pairing (see \cite{Broglia:73} and refs. therein). In the other extreme, that is, in systems displaying an extreme neutron excess like $^{11}$Li, this interplay has been found to be the basic source of stability of the $|\text{gs}(^{11}\text{Li})\rangle$ state. 
Within this context, we remind of the fact that pairing in nuclei is, as a rule, and in keeping with the paper that started the subject\cite{Bohr:58}, related to the presence of a gap $\Delta(=G\alpha'_0)$, in the low lying spectrum. Thus, eventually to the concept of odd--even mass difference (OEMD), (within this context see the chapter contributed by Brown to this Volume), quantity which provides information concerning  the $A$--dependence of the gap and of the pairing coupling constant ($\Delta\approx 12$MeV$/\sqrt{A}, G\approx 25$MeV$/A$), see \cite{Bohr:75,Bohr:69} and refs therein; it is of notice that the $A$--dependence of the nuclear pairing gap is still an open question. Within this context cf. Fig. 10.10, p. 236 of ref. \cite{Brink:05} .

Now, $^9_3$Li$_6$ and  $^{11}_3$Li$_8$ are bound systems, while  $^{10}_3$Li$_7$ is not. How does one proceed in such a case concerning the OEMD? Arguably dynamically, through a bootstrap mechanism of pair correlation induced by the exchange, between Cooper pair partners, of exotic modes of nuclear excitation (dipole pigmy resonances), arising from the sloshing back and forth of the unbound halo neutron field, with respect to the  $^9$Li proton core.

This is a possible new mechanism to break gauge symmetry (within this context cf. L. N. Cooper contribution to the volume BCS: 50 years, World Scientific, Singapore (2011)p.18) likely to be found in weakly bound nuclei displaying $s$-- and $p$-- levels near threshold, connected with the fact that in e.g. the case of $^{11}$Li, the correlation length $\xi=\hbar v_F/2 E_{corr}\approx 20$fm ($v_F\approx 0.1 c, E_{corr}\approx 0.5$ MeV). While the bare $^1S_0$ interaction ($G_{bare}$) can be viewed, essentially as a contact interaction,thus receiving small, coherent, contributions from many $l$--multipolarities in particular high values of $l$ and thus little operative in the present circumstance.
The exchange of low lying, collective vibrations like the pigmy resonance ($\approx 2$MeV in $^{11}$Li), provides correlation over distances of the order of the coherence length ($G_{ind}$). In keeping with the fact that in FMBS like the nucleus, Cooper condensation requires a critical value of the pairing coupling constant $G_{cr}$, the bootstrap nuclear pair correlation mechanism, intimately related with a long coherence length, reminds some of the concepts which are at the basis of the Josephson effect. Halo pairing vibrations, of which the $|gs(^{11}$Li)$\rangle$ state is a concrete embodiment, are likely to be a new elementary mode of excitation, arising from a (dynamical) breaking of gauge symmetry. Because they are expected to be weakly correlated, extended (low--$k$) fragile nuclear objects (mainly based on $s$-- and $p$-- states at threshold\footnote{Within this context, it is of notice that the associated, mean field, antipairing effect discussed in refs. \cite{Hamamoto:03}, and intimately related with the bare (contact--like) interaction is, arguably, overwhelmed by (dynamical) medium polarization effects (see also \cite{Hamamoto:04} and \cite{Bennaceur:00})}), it is likely that their properties, e.g. their intrinsic two--particle transfer cross section, can be altered when shifted around, e.g. as excited states in nuclear species with different proton number. In any case, their most distinct feature, namely that of displaying a pigmy resonance at a relative excitation energy of few MeV, necessary although not sufficient condition for this new mode to exist, can be instrumental in their  quest.  Within this context, one could expect to find this halo, pair addition mode, as an excited $0^+$ in $^{12}$Be.


\subsection{Hindsight}
Reaching to the limits of stability associated with light drip line nuclei, and to situations in which medium polarization and spatial quantization effects become overwhelming, one is confronted with elementary modes of nuclear excitation in which, dynamic, fluctuation effects are as important as, static, mean field effects. NFT within the Bloch--Horowitz (Dyson) set up which allows to sum to infinite order little convergent processes, seems to be able to provide an accurate description of these systems which are also predictive. From these studies it perspires that fragile objects like $|\text{gs}(^{11}\text{Li})\rangle$ ($S_{2n}\approx 400$ MeV), the halo pair addition mode of the $N=6$ closed shell system $^9$Li, may be resilient and flexible enough to be transported, as elementary modes of excitation from one nuclear species to another one, without the risk to break the Cooper pair. This  halo vibrational mode may be directly observed in $L=0$, two--particle transfer reactions to excited states, or in terms of E1 decay of eventual pigmy resonances built on top of it. It is not unthinkable that  at the light end of the periodic table, but not only, nuclear species have played this game, namely to build, phonon mediated, pairing correlations based on $s$-- and $p$-- almost bound, virtual single--particle states to reach the limits of the drip lines.

\section{Conclusions}
Nuclear structure and reactions are but just two aspects of the same physics, mainly connected with bound and continuum states respectively, a difference which becomes further blurred in the description of exotic nuclei at large and of halo, neutron drip line nuclei in particular. Mostly so, in the case of BCS--pairing like models and two--nucleon transfer reactions.

The validity of the original BCS description of the intrinsic state of pairing rotational bands (coherent state) and the QRPA restoration of gauge symmetry associated with the AGN mode,together with the development of the tools needed to describe the interplay between successive and simultaneous transfer, taking into account also the non--orthogonality of the wavefunctions describing the motion of nucleons in target and projectile, lead to absolute two--particle transfer differential cross sections which reproduce the experimental data well below the 10\% level of accuracy. It is of notice that this development is to be compared,  with all the caveats of the case, to that associated with Josephson's application of BCS theory to Cooper pair tunnelling experiments and the associated analysis based on Cohen--Falicov--Phillips pair tunnelling description.

While in a FMBS like the atomic nucleus supercurrents are not possible, and phase transitions (normal--superfluid) are blurred due to strong fluctuations in gauge space, one can study the emergence of these phenomena in terms of individual quantal states. Furthermore, investigate the many phonon pairing -vibrational-  spectrum and its melting into a Cooper pair condensate, as well as search for new, exotic embodiments of dynamical spontaneous symmetry breaking of gauge invariance, like e.g. halo pairing vibrational modes. By properly adjusting the bombarding conditions and the selection of the superfluid nuclei to study in the transfer process, one can, through $Q$--value effects, open or close individual transfer channels, e.g. one--particle transfer channels. In this way the relative importance of simultaneous and successive transfer will be shifted in favour of the first one, eventually allowing to study it in detail. To which extent one could observe pairing interaction mediated transfer is an open question. In any case, its search may provide a fresh, closer look, to Bardeen's arguments concerning superfluid tunnelling.

The possibility of studying the effect temperature and  rotation (similar to magnetic field effects in condensed matter) have on nuclear pairing, and thus in particular on the $\omega$--dependent part of the pairing interaction arising from the exchange of collective vibrations between the partners of Cooper pairs, can shed new insight into BCS--like pairing in fermionic system. This is in keeping with the fact that one will be able to study the normal--superfluid phase transition not only in terms of phononic density of states changes as a function of $T$ and $\omega$, but also in terms of the melting of the phonon collectivity associated with these changes. This is in keeping with the fact that in nuclei we are in a situation in which temperature and rotations have energy scales similar to that of the correlation energy determining the collectivity of vibrational modes (phonons). In other words, and making again use of the analogy with the case of superconductivity in metals, one seems to be  confronted, in the nuclear case, with situations in which temperature and magnetic fields lead, among other things, to a melting or at least to a strong perturbation of the crystal lattice, and thus of the associated phonon spectrum.

Discussions an collaboration with F. Barranco, B. Bayman, A. Idini and E. Vigezzi are gratefully acknowledged. We want also to thank L. Zetta and P. Guazzoni for clarification and access regarding $(p,t)$ data. The research work of GP was supported by the Ministry of Science and Innovation of Spain, grant FPA2009--07653.

\bibliographystyle{ws-rv-van}

\end{document}